\documentclass[preprint,preprintnumbers,nofootinbib,amsmath,amssymb]{revtex4-2}
\usepackage{amssymb,latexsym,amsmath}

\usepackage[activeacute,english]{babel}
\usepackage{fancyhdr}
\usepackage{textcomp}
\usepackage{amsfonts}
\usepackage{graphicx}
\usepackage{graphics}
\usepackage{subfig}

\usepackage{xcolor}

\newcommand{\pa}{\partial}

\newcommand{\om}{\omega}

\newcommand{\rar}{\rightarrow}

\def\be {\begin {equation}}
\def\ee {\end {equation}}

\newcommand{\ba}{\begin{array}}
\newcommand{\ea}{\end{array}}
\newcommand{\bea}{\begin{eqnarray}}
\newcommand{\eea}{\end{eqnarray}}
\newcommand{\bi}{\begin{itemize}}
\newcommand{\ei}{\end{itemize}}

\begin{document}

\title{Classical $n$-body system in volume variables. II. Four-body case}

\author{A. M. Escobar-Ruiz}
\email{admau@xanum.uam.mx}
\affiliation{
Departamento de F\'isica,
Universidad Aut\'onoma Metropolitana-Iztapalapa, San Rafael Atlixco 186,
M\'exico, CDMX, 09340 M\'exico}

\author{Alexander V Turbiner}
\email{turbiner@nucleares.unam.mx}
\affiliation{Instituto de Ciencias Nucleares, UNAM, M\'exico DF 04510, Mexico}

\begin{abstract}
{It is evident that the positions of 4 bodies in $d>2$ dimensional space can be identified with vertices of a tetrahedron. Square of volume of the tetrahedron, weighted sum of squared areas of four facets and weighted sum of squared edges are called the volume variables. A family of translation-invariant potentials which depend on volume variables alone is considered as well as solutions of the Newton equations which solely depend on volume variables. For the case of zero angular momentum $L=0$ the corresponding Hamiltonian, which describes these solutions, is derived. Three examples are studied in detail: (I) the (super)integrable 4-body closed chain 
of harmonic oscillators for $d>2$ (the harmonic molecule), (II) a generic, two volume variable dependent potential whose trajectories possess a constant moment of inertia ($d>1$), and (III) the 4-body anharmonic oscillator for $d \geq 1$.
This work is the second of the sequel: the first one [IJMPA 36, No. 18 (2021)] was dedicated to study the
3-body classical problem in volume variables.
}

\bigskip

\end{abstract}

\keywords{Classical mechanics, 4-body system,symmetric reduction, harmonic pairwise interaction}

\maketitle

\section{Introduction}
\label{Introduction}

For two interacting point-like particles of masses $m_1$ and $m_2$ in $\mathbb{R}^d$ ($d>1$), with scalar potential $V=V(r_{12})$ that depends on the relative distance $r_{12}=\mid {\bf r}_1-{\bf r}_2\mid$, the Lagrangian is given by
\begin{equation}
\label{H2bo}
  {\cal L}_2 \ = \ \frac{1}{2}m_1\, {\dot{{\bf r}}_1}^{2}\ + \ \frac{1}{2}m_2\, {\dot{{\bf r}}_2}^{2} \ - \ V(r_{12}) \ ,
\end{equation}
here ${\bf r}_i\in \mathbb{R}^d$ is the position vector of the $i$th particle and $\dot{{\bf r}}_i \equiv \frac{d}{dt}{\bf r}_i $ is the velocity, $i=1,2$. The $d$-dimensional center-of-mass (cms) motion can be separated out, and we arrive at the $d$-dimensional problem in the \emph{space of relative motion}. In the case of \emph{zero total angular momentum} the $d$-dimensional 2-body dynamics, which the Lagrangian (\ref{H2bo}) governs, is reduced to one-dimensional dynamics in variable $r_{12}$, governed by the Lagrangian
\begin{equation}
\label{H2bored}
  \tilde {\cal L}_2 \ = \ \frac{1}{2}\mu\, {\dot r}_{12}^{2}  \ - \ V(r_{12}) \ ,
\end{equation}
where $\mu=\frac{m_1\,m_2}{m_1+m_2}$ stands for the reduced mass. The symmetry reduction ${\cal L}_2 \rightarrow \tilde {\cal L}_2$ {from $(2d)$-dimensional coordinate space to one-dimensional space of relative radial motion} is well described in textbooks on Classical Mechanics, see e.g. \cite{LandauLifshitzBook}. We should only emphasize that the \emph{generalized coordinate} $r_{12}$ in (\ref{H2bored}) plays a fundamental role in this symmetry reduction, it possesses a clear geometrical meaning: it corresponds to the length of the interval which connects two point masses. This variable can be naturally called the \emph{interval (length) of interaction}. Formal $Z_2$ invariance $r_{12} \rar -r_{12}$ of the kinetic energy suggests to introduce the
{\it volume variable} $\rho_{12}=r_{12}^2$.

\bigskip

For the three-body case in $\mathbb{R}^d$ ($d>1$) with three arbitrary masses, the emergent geometrical object is the \textit{triangle of interaction} formed by the body positions as vertices. The lengths of its edges are the 3 relatives distances $r_{ij}=\mid {\bf r}_i-{\bf r}_j\mid$ between particles. Following the proposal by
J.L.~Lagrange \cite{Lagrange}, in our previous paper \cite{IJMP1} it was shown that the classical dynamics of the system at zero total angular momentum is reduced to the dynamics of the triangle with fixed baricenter: it is described by a reduced Hamiltonian ${\cal H}_{L=0}$ with three degrees of freedom $r_{12}, r_{13}, r_{23}$,
see also \cite{Murnaghan, Kampen}. It was called the $r-$representation {of 3-body at zero angular momentum}, where the variables $r_{ij}$ (complemented by their corresponding canonical momenta) are used. As a realization of the formal $Z_2$ invariance of the Hamiltonian ${\cal H}_0$: $r_{ij} \rar -r_{ij}$ in \cite{IJMP1} it was proposed to introduce another set of generalized coordinates $\rho_{ij}=r_{ij}^2$.
It was called the $\rho-$representation of ${\cal H}_{L=0}$. As a next step, for the particular case
of three equal masses $m_1=m_2=m_3 \equiv m$, the lowest order symmetric polynomial invariants
\begin{equation}
\label{sigv}
\begin{aligned}
 & z_1 \ = \ \rho_{12} \  + \ \rho_{23} \  + \ \rho_{31} \  ,
\\
 & z_2 \ = \ \rho_{12}\,\rho_{23} \  + \ \rho_{12}\,\rho_{31} \ + \ \rho_{23}\,\rho_{31} \  ,
\\
 & z_3 \ = \ \rho_{12} \, \rho_{23} \, \rho_{31} \  .
\end{aligned}
\end{equation}
were proposed to use as new generalized coordinates, they were called the {\it geometrical variables}. These coordinates encode the entire discrete symmetry $\mathbb{Z}_2^{\otimes3} \oplus {S}_3 \oplus {S}_3$  of the kinetic energy in ${\cal H}_0$, they are simultaneously invariant under the reflections $r_{ij} \rightarrow -r_{ij}$, the permutations of the bodies and the permutation of the edges of the triangle of interaction. Moreover, for potentials that depends on first two variables only in (\ref{sigv}), $V=V(z_1, z_2)$, {which are the volume variables for 3-body problem}, a further reduction to two degrees of freedom was made and the effective (reduced) Hamiltonian was constructed as well. Note that the volume variable ${\cal S}_2=(4 z_2-z_1^2)/16$ is the square of the \textit{area} of the triangle of interaction whereas the volume variable ${\cal P}_2=z_1$ is sum of its three edges squared. The main properties of the aforementioned representations were illustrated in \cite{IJMP1} by several examples explicitly such as the three-body chain of harmonic oscillators ($d>2$), the celebrated three-body Newton problem in Celestial Mechanics and the three-body choreography in
$d = 2$ on the algebraic lemniscate.

\bigskip

In the case of four bodies with $d$-degrees of freedom ($d > 2$), the six relative distances $r_{ij}$ are nothing but the edges of the {\it tetrahedron of interaction} which is formed by taking the bodies positions as vertices. For arbitrary potentials $V=V(r_{ij})$, if assuming zero total angular momentum, a reduced Hamiltonian {${\cal H}_{\rm red}\equiv {\cal H}_{L=0}$} in the space of relative motion has six degrees of freedom, it will be constructed in both $r-$ and $\rho-$representations. Interestingly, it describes a six-dimensional classical particle moving in a curved space, (see also \cite{4bodyQ} for the corresponding quantum problem). It is worth mentioning that {as early as} in 1900 O. Dziobek \cite{Dziobek} already considered the central configurations in $n$-body Newtonian gravity in the $r-$representation. As explained before, this representation employs elements of dimension one (edges of the polytope) as generalized coordinates: it continues to be used, attracting the attention in contemporary Celestial Mechanics (see Refs.\cite{Santoprete, Cors, Albouy, Saari} and references therein). Here, we explore alternative generalized variables of different dimensionalities.

\bigskip

The goal of the present study is to construct the {six (geometrical) variables} that encode the discrete symmetries of the tetrahedron of interaction, {then build three volume variables out of those and for which the kinetic energy of the effective (reduced) free Hamiltonian is a polynomial function in the phase space. We will consider a (sub)-family of potentials which depend on the volume variables alone, which includes many physically important particular cases, three of those will be studied explicitly.} The novelty of the present approach is the explicit introduction of elements of different dimensionality of the tetrahedron of interaction as generalized coordinates. By doing so, an interesting link between the dynamics of the system and the theory of polytopes occurs. {It realizes a symmetry reduction from $(4d)$-dimensional coordinate space to six-dimensional space.}

\bigskip

\section{Four-body system: case $d>2$}
\label{planar case}

\bigskip

We consider for $d>2$ a classical system of four interacting point-like particles with masses $m_1,m_2,m_3$ and $m_4$, respectively. The non-relativistic Lagrangian is of the form,
\begin{equation}
\label{H}
   {\cal L}\ =\ {\cal T} \ - \ V(r_{12},\,r_{13},\,r_{14},\,r_{23},\,r_{24},\,r_{34})  \ ,
\end{equation}
where the kinetic energy is given by
\begin{equation}
\label{Tflat}
   {\cal T}\ =\ \frac{1}{2}m_1\, {\dot{{\bf r}}_1}^{2}\ + \ \frac{1}{2}m_2\, {\dot{{\bf r}}_2}^{2}\ + \ \frac{1}{2}m_3\, {\dot{{\bf r}}_3}^{2}\ + \ \frac{1}{2}m_4\, {\dot{{\bf r}}_4}^{2}\ ,
\end{equation}
here ${\bf r}_i \in \mathbb{R}^d$ is the vector position of the $i$th body, $\dot{{\bf r}_i} \equiv \frac{d}{dt}{\bf r}_i $ and the scalar potential $V$ depends on the relative distances
\[
r_{ij} \ \equiv \  \mid {\bf r}_i-{\bf r}_j\mid \ ,
\]
between particles only. This implies that ${\cal L}$ is rotationally symmetric and translationally invariant. The configuration space is $4d$-dimensional $\mathbb{R}^d({\bf r}_1) \times \mathbb{R}^d({\bf r}_2) \times\mathbb{R}^d({\bf r}_3)\times\mathbb{R}^d({\bf r}_4)$.

Equivalently, the Hamiltonian of the system reads

\begin{equation}
\label{Hamil}
   {\cal H}\ =\ \frac{{{\bf p}}_1^{2}}{2\,m_1}\ + \ \frac{{{\bf p}}_2^{2}}{2\,m_2}\ + \ \frac{{{\bf p}}_3^{2}}{2\,m_3}\ + \ \frac{{{\bf p}}_4^{2}}{2\,m_4} \ + \ V(r_{12},\,r_{13},\,r_{14},\,r_{23},\,r_{24},\,r_{34})  \ ,
\end{equation}

with ${{\bf p}}_j =\frac{\pa\,{\cal L}}{\pa \dot{{\bf r}}_j}  $, $j=1,2,3,4$, being the canonical momentum vectorial variables, respectively. Hence, the phase space is $8d$-dimensional.
Since the center-of-mass motion can be separated completely, the internal relative motion of the system is characterized by $3d$ degrees of freedom. In this space of relative motion, let be
\begin{equation}
\{q_1,\,q_2,\,q_3,\ldots,\,q_{3d}\} \ ,
\end{equation}
any set of generalized coordinates. For instance, one possibility is to take the three $d$-dimensional vectorial Jacobi coordinates defined by the linear relations

\begin{equation}
\label{Jacobi}
\begin{aligned}
    & {\bf r}^{(J)}_{1} \ = \ \sqrt{\frac{m_1\,m_2}{m_1+m_2}}\left({\bf r}_{2}\ - \ {\bf r}_{1}\right)
\\ &
{\bf r}^{(J)}_{2} \ = \ \sqrt{\frac{(m_1+m_2)\,m_3}{m_1+m_2+m_3}}\left({\bf r}_{3}\ - \ \frac{m_1\,{\bf r}_{1}+m_2\,{\bf r}_{2}}{m_1+m_2}\right)
\\ &
{\bf r}^{(J)}_{3} \ = \ \sqrt{\frac{(m_1+m_2+m_3)\,m_4}{m_1+m_2+m_3+m_4}}\left({\bf r}_{4}\ - \ \frac{m_1\,{\bf r}_{1}+m_2\,{\bf r}_{2}+m_3\,{\bf r}_{3}}{m_1+m_2+m_3}\right)
     \  .
\end{aligned}
\end{equation}
In these coordinates the $3d$-dimensional kinetic energy term ${\cal T}_{\rm rel}^{(3d)}$ of the relative motion becomes diagonal. Explicitly,
\begin{equation}
\label{Tflat-diag}
   2\,{\cal T}\ = \ {\cal T}_{\rm center-of-mass}^{(d)} \ + \ {\cal T}_{\rm rel}^{(3d)} \ = \ {({{\bf P}}_{{}_{\bf Y}})}^{2} \ + \ {({\bf p}^{(J)}_{1})}^2 \ + \ {({\bf p}^{(J)}_{2})}^2\ + \ {({\bf p}^{(J)}_{3})}^2 \ .
\end{equation}
The kinetic energy of relative motion is the sum of kinetic energies in the Jacobi coordinate directions. This property will be exploited in the study of the 4-body harmonic chain.

For future convenience, we select the center-of-mass system as the inertial frame, namely, ${\bf Y} =0,\, {{{\bf P}}_{{}_{\rm Y}}=M\,{\dot{\bf Y}}=0}$. Here
\[
{\bf Y}\ \equiv \ \frac{m_1\,{\bf r}_1  +  m_2\,{\bf r}_2  +  m_3\,{\bf r}_3+  m_4\,{\bf r}_4}{M} \quad , \qquad M=m_1+m_2+m_3+m_4 \ .
\]
Accordingly, from the original $4d-$dimensional problem we obtain a reduced Hamiltonian, ${\cal H} \rightarrow {\cal H}_{\rm rel}$, in the space of relative motion

\begin{equation}
\label{HrelM}
{\cal H}_{\rm rel} \ = \ {({\bf p}^{(J)}_{1})}^2 \ + \ {({\bf p}^{(J)}_{2})}^2\ + \ {({\bf p}^{(J)}_{3})}^2 \ + \  V({\bf r}^{(J)}_{1},\,{\bf r}^{(J)}_{2},\,{\bf r}^{(J)}_{3})   \ .
\end{equation}

\section{Reduced Hamiltonian at zero angular momentum}

\subsection{$r$-representation}

Let us focus on the description of the trajectories that solely depend on the relative distances $r_{ij}=\mid {\bf r}_i - {\bf r}_j\mid$ and possess zero total angular momentum. The phase space associated to the relative motion is $6d-$dimensional. Let us denote it by $\Gamma_{\rm rel}$. Now, in the space of relative motion parametrized by Jacobi-vectorial variables we introduce a further change of coordinates
\begin{equation}
\label{CC4b}
\{{\bf r}^{(J)}_{1},\,{\bf r}^{(J)}_{2},\,{\bf r}^{(J)}_{3}\}\ \Rightarrow \
\{  r_{12},\,r_{13},\,r_{14},\,r_{23},\,r_{24},\,r_{34},\,\phi_1,\,\phi_2,\,\ldots ,\,\phi_{(3d-6)}  \} \ ,
\end{equation}

where the collection \{$\phi_i$\}, $i=1,2,3,\ldots,(3d-6)$, stands for any set of variables such that the transformation (\ref{CC4b}) is canonical.  For the potential $V$ does not involve the $\phi-$variables, it follows that the hypersurface
\begin{equation}
\label{IC4b}
\Gamma_r \ \equiv \ \Gamma_{\rm rel}|_{\ p_{_{\phi_1}} =\,p_{_{\phi_2}}=\,p_{_{\phi_3}}=\,\ldots\,= \,p_{_{\phi_{(3d-6)}}}=0} \quad ; \,\qquad  \Gamma_r \ \subset \ \Gamma_{\rm rel}
\end{equation}
is an invariant manifold\cite{Wiggins}. In other words, any trajectory of ${\cal H}_{\rm rel}$ (\ref{HrelM}) with initial conditions $\ p_{_{\phi_1}} =\,p_{_{\phi_2}}=\,p_{_{\phi_3}}=\,\ldots\,= \,p_{_{\phi_{(3d-6)}}}=0$ will remain in $\Gamma_r$ under time evolution. It should be noted that the $\phi-$variables are not cyclic coordinates. In general, they will appear in the kinetic energy term ${\cal T}_{\rm rel}^{(3d)}$ explicitly. Hence, the associated momenta $p_{\phi}$ are not first Liouville integrals but \textit{particular integrals} of motion\cite{Turbiner:2013p}. They are conserved quantities for specific initial conditions only.

The trajectories possessing zero total angular momentum necessarily lie in $\Gamma_r$. Therefore, by taking the restriction of ${\cal H}_{\rm rel}$ on $\Gamma_r$, we form an effective (reduced) Hamiltonian (${\cal H}_{\rm rel} \rightarrow H_r$)
\begin{equation}
\label{Hredzeroam}
\begin{aligned}
  {H}_{r} \  & = \ \frac{1}{2}\bigg[\,\frac{p_{12}^2}{m_{12}} \ + \ \frac{p_{13}^2}{m_{13}} \ + \  \frac{p_{14}^2}{m_{14}} \ + \   \frac{p_{23}^2}{m_{23}} \ + \  \frac{p_{24}^2}{m_{24}} \ + \ \frac{p_{34}^2}{m_{34}} \ + \ \frac{r_{12}^2+r_{13}^2-r_{23}^2}{m_1\,r_{12}\,r_{13}}\,p_{12}\,p_{13}
\\ &
\qquad   \ +
  \ \frac{r_{12}^2+r_{23}^2-r_{13}^2}{m_2\,r_{12}\,r_{23}}\,p_{12}\,p_{23}\ + \
\frac{r_{23}^2+r_{13}^2-r_{12}^2}{m_3\,r_{23}\,r_{13}}\,p_{23}\,p_{13}
\ + \
\frac{r_{24}^2+r_{14}^2-r_{12}^2}{m_4\,r_{24}\,r_{14}}\,p_{24}\,p_{14}
\\ &
\qquad   \ + \
   \frac{r_{12}^2+r_{14}^2-r_{24}^2}{m_1\,r_{12}\,r_{14}}\,p_{12}\,p_{14} \ + \
   \frac{r_{12}^2+r_{24}^2-r_{14}^2}{m_2\,r_{12}\,r_{24}}\,p_{12}\,p_{24}\ + \
   \frac{r_{13}^2+r_{34}^2-r_{14}^2}{m_3\,r_{13}\,r_{34}}\,p_{13}\,p_{34}
   \\ &
\qquad   \ + \
   \frac{r_{14}^2+r_{34}^2-r_{13}^2}{m_4\,r_{14}\,r_{34}}\,p_{14}\,p_{34}  \ + \
   \frac{r_{13}^2+r_{14}^2-r_{34}^2}{m_1\,r_{13}\,r_{14}}\,p_{13}\,p_{14} \ + \
   \frac{r_{23}^2+r_{24}^2-r_{34}^2}{m_2\,r_{23}\,r_{24}}\,p_{23}\,p_{24}
   \\ &
\qquad   \ + \
   \frac{r_{23}^2+r_{34}^2-r_{24}^2}{m_3\,r_{23}\,r_{34}}\,p_{23}\,p_{34} \ + \
   \frac{r_{24}^2+r_{34}^2-r_{23}^2}{m_4\,r_{24}\,r_{34}}\,p_{24}\,p_{34}   \,\bigg] \ + \  V     \ ,
\end{aligned}
\end{equation}
here $p_{ij}$ is the canonical momentum associated with $r_{ij}$.
The expression (\ref{Hredzeroam}) we call the $r-$representation of the reduced Hamiltonian. 
The kinetic energy term (equivalently, the free Hamiltonian) is a polynomial function 
of the momentum variables with rational $r-$dependent coefficients.

\begin{figure}[htp]
  \centering
\includegraphics[width=13.0cm]{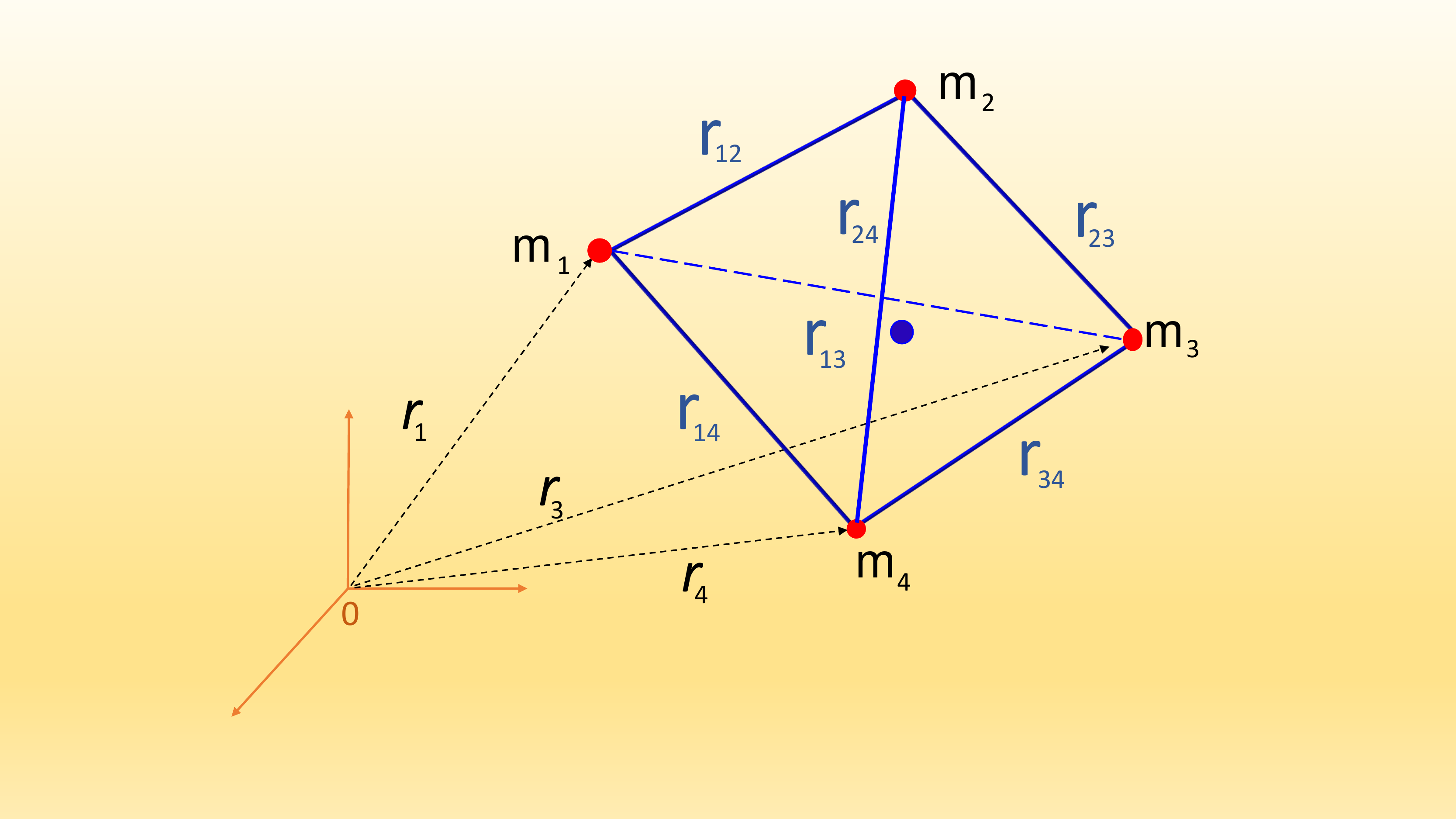}
  \caption{Tetrahedron of interaction in $d=3$: the individual coordinate vectors ${\bf r}_i$ mark positions of vertices of the tetrahedron with edges $r_{ij}$. Accordingly, its four faces are triangles. The center-of-mass is marked by a (blue) bubble.}
\label{Fig3}
\end{figure}

\subsubsection{The metric $g^{\mu \nu}(r)$}

In (\ref{Hredzeroam}), we identify the components of the associated cometric $g^{\mu \nu}(r)$ with the coefficients in front of the quadratic terms in momentum variables $(p_{12},\,p_{31},\,p_{23})$. Explicitly,
\begin{equation}
\label{gmn33-red}
 g^{\mu \nu}(r) =
      \left(
\begin{array}{cccccc}
 \frac{m_1+m_2}{2 m_1 m_2} & \frac{r_{12}^2+r_{13}^2-r_{23}^2}{4 m_1 r_{12} r_{13}} & \frac{r_{12}^2+r_{14}^2-r_{24}^2}{4 m_1 r_{12} r_{14}} & \frac{r_{12}^2-r_{13}^2+r_{23}^2}{4 m_2 r_{12} r_{23}} & \frac{r_{12}^2-r_{14}^2+r_{24}^2}{4 m_2 r_{12} r_{24}} & 0 \\
 \frac{r_{12}^2+r_{13}^2-r_{23}^2}{4 m_1 r_{12} r_{13}} & \frac{m_1+m_3}{2 m_1 m_3} & \frac{r_{13}^2+r_{14}^2-r_{34}^2}{4 m_1 r_{13} r_{14}} & \frac{-r_{12}^2+r_{13}^2+r_{23}^2}{4 m_3 r_{13} r_{23}} & 0 & \frac{r_{13}^2-r_{14}^2+r_{34}^2}{4 m_3 r_{13} r_{34}} \\
 \frac{r_{12}^2+r_{14}^2-r_{24}^2}{4 m_1 r_{12} r_{14}} & \frac{r_{13}^2+r_{14}^2-r_{34}^2}{4 m_1 r_{13} r_{14}} & \frac{m_1+m_4}{2 m_1 m_4} & 0 & \frac{-r_{12}^2+r_{14}^2+r_{24}^2}{4 m_4 r_{14} r_{24}} & \frac{-r_{13}^2+r_{14}^2+r_{34}^2}{4 m_4 r_{14} r_{34}} \\
 \frac{r_{12}^2-r_{13}^2+r_{23}^2}{4 m_2 r_{12} r_{23}} & \frac{-r_{12}^2+r_{13}^2+r_{23}^2}{4 m_3 r_{13} r_{23}} & 0 & \frac{m_2+m_3}{2 m_2 m_3} & \frac{r_{23}^2+r_{24}^2-r_{34}^2}{4 m_2 r_{23} r_{24}} & \frac{r_{23}^2-r_{24}^2+r_{34}^2}{4 m_3 r_{23} r_{34}} \\
 \frac{r_{12}^2-r_{14}^2+r_{24}^2}{4 m_2 r_{12} r_{24}} & 0 & \frac{-r_{12}^2+r_{14}^2+r_{24}^2}{4 m_4 r_{14} r_{24}} & \frac{r_{23}^2+r_{24}^2-r_{34}^2}{4 m_2 r_{23} r_{24}} & \frac{m_2+m_4}{2 m_2 m_4} & \frac{-r_{23}^2+r_{24}^2+r_{34}^2}{4 m_4 r_{24} r_{34}} \\
 0 & \frac{r_{13}^2-r_{14}^2+r_{34}^2}{4 m_3 r_{13} r_{34}} & \frac{-r_{13}^2+r_{14}^2+r_{34}^2}{4 m_4 r_{14} r_{34}} & \frac{r_{23}^2-r_{24}^2+r_{34}^2}{4 m_3 r_{23} r_{34}} & \frac{-r_{23}^2+r_{24}^2+r_{34}^2}{4 m_4 r_{24} r_{34}} & \frac{m_3+m_4}{2 m_3 m_4} \\
\end{array}
\right)  \ .
\end{equation}
Its determinant $\mathcal{D}_m = \text{Det}[g^{\mu \nu}(r)]$ possesses the remarkable factorization property
\begin{equation}
\label{det-m}
  \mathcal{D}_m \ = \ \frac{9\,{(m_1+m_2+m_3+m_4)}^2}{64\ m_1^3\,m_2^3\,m_3^3\,{m_4^3}\ r_{12}^2\,r_{13}^2\,r_{14}^2\,r_{23}^2\,r_{24}^2\,r_{34}^2}\, {\cal V} \,(\,{\cal S}\,{\cal P} \ - \ 144\,m_1\,m_2\,m_3\,m_4\,{\cal V}\,) \ ,
\end{equation}
where
\begin{equation}
\label{VVV}
\begin{aligned}
{\cal V} \ & = \ \frac{1}{288} \bigg[\,-2 r_{34}^2 r_{12}^4-2 r_{34}^4 r_{12}^2-2 r_{13}^2 r_{23}^2 r_{12}^2+2 r_{14}^2 r_{23}^2 r_{12}^2
+2 r_{13}^2 r_{24}^2 r_{12}^2-2 r_{14}^2 r_{24}^2 r_{12}^2 \ + \
\\ &
2 r_{13}^2 r_{34}^2 r_{12}^2+2 r_{14}^2 r_{34}^2 r_{12}^2+2 r_{23}^2 r_{34}^2 r_{12}^2+2 r_{24}^2 r_{34}^2 r_{12}^2-2 r_{14}^2 r_{23}^4-2 r_{13}^2 r_{24}^4-2 r_{14}^4 r_{23}^2 \ + \
\\ &
2 r_{13}^2 r_{14}^2 r_{23}^2-2 r_{13}^4 r_{24}^2
+2 r_{13}^2 r_{14}^2 r_{24}^2+2 r_{13}^2 r_{23}^2 r_{24}^2+2 r_{14}^2 r_{23}^2 r_{24}^2-2 r_{13}^2 r_{14}^2 r_{34}^2+2 r_{14}^2 r_{23}^2 r_{34}^2
\\ &
\ + \ 2 r_{13}^2 r_{24}^2 r_{34}^2\ - \ 2 r_{23}^2 r_{24}^2 r_{34}^2\,\bigg] \ ,
\end{aligned}
\end{equation}

\begin{equation}
\label{VVS}
\begin{aligned}
{\cal S} \  = \ & m_1\,m_2\,m_3\,\big(2 r_{13}^2 r_{12}^2+2 r_{23}^2 r_{12}^2+2 r_{13}^2 r_{23}^2-r_{13}^4-r_{23}^4-r_{12}^4  \big) \ + \
\\ &
m_1\,m_2\,m_4\,\big( 2 r_{14}^2 r_{12}^2+2r_{24}^2 r_{12}^2+2 r_{14}^2 r_{24}^2-r_{12}^4-r_{14}^4-r_{24}^4\big) \ + \
\\ &
m_1\,m_3\,m_4\,\big(2 r_{14}^2 r_{13}^2+2 r_{34}^2 r_{13}^2+2 r_{14}^2 r_{34}^2-r_{14}^4-r_{34}^4 -r_{13}^4\big) \ +\
\\ &
m_2\,m_3\,m_4\,\big( 2 r_{24}^2 r_{23}^2+2 r_{34}^2 r_{23}^2+2 r_{24}^2 r_{34}^2-r_{24}^4-r_{34}^4-r_{23}^4 \big) \ ,
\end{aligned}
\end{equation}
\\
\begin{equation}
\label{VVP}
{\cal P} \ = \ \frac{m_1 m_2 r_{12}^2+m_1 m_3 r_{13}^2+m_1 m_4 r_{14}^2+m_2 m_3 r_{23}^2+m_2 m_4 r_{24}^2+m_3 m_4 r_{34}^2}{m_1+m_2+m_3+m_4} \ ,
\end{equation}
are called volume variables. The variable $\cal V$ is the square of the volume of the tetrahedron of interaction, $\cal S$ is the weighted sum of the 4 areas (squared) of the faces (triangles) of the tetrahedron whilst $\cal P$ is the weighted sum of the 6 edges (squared) of the tetrahedron. In fact, $\cal P$ coincides with the moment of inertia of the system.

\noindent
\textbf{Remark.} The above determinant $\mathcal{D}_m$ (\ref{det-m}) is a rational function in $r$-variables. It effectively depends on the four coordinates $\cal V$, $\cal S$, $\cal P$ and $T \equiv r_{12}^2\,r_{13}^2\,r_{14}^2\,r_{23}^2\,r_{24}^2\,r_{35}^2$. Note that it vanishes, $\mathcal{D}_m=0$, when i) the volume of the tetrahedron of interaction is equal to zero, ii) the tetrahedron is regular, or iii) a triple-body collision occurs. It is singular at the two-body collision point.

\subsection{$\rho$-representation}

The cometric (\ref{gmn33-red}) is invariant formally under reflections $\mathbb{Z}_2\oplus \mathbb{Z}_2\oplus \mathbb{Z}_2\oplus \mathbb{Z}_2\oplus \mathbb{Z}_2\oplus \mathbb{Z}_2$,
\[
r_{ij}\, \rightarrow \, - r_{ij} \qquad \ i,j=1,2,3,4 \quad (i\neq j) \ .
\]
If we introduce new variables,
\begin{equation}\label{rhovar}
\rho_{ij}\ = \  r_{ij}^2 \ ,
\end{equation}
with the corresponding canonical momenta
\begin{equation}\label{prhovar}
P_{ij}\, = \,  \frac{1}{2\,r_{ij}}\,p_{ij}  \ ,
\end{equation}
we immediately arrive at the ${\mathbb{Z}}_2$-symmetry reduced Hamiltonian
\begin{equation}
\label{Hredrho}
\begin{aligned}
  {H}_{\rho} \  & = \\ & \ 2\,\bigg[\,\frac{\rho_{12}\,P_{12}^2}{m_{12}} \ + \ \frac{\rho_{13}\,P_{13}^2}{m_{13}} \ + \  \frac{\rho_{14}\,P_{14}^2}{m_{14}} \ + \   \frac{\rho_{23}\,P_{23}^2}{m_{23}} \ + \  \frac{\rho_{24}\,P_{24}^2}{m_{24}} \ + \ \frac{\rho_{34}\,P_{34}^2}{m_{34}} \ + \ \frac{\rho_{12}+\rho_{13}-\rho_{23}}{m_1}\,P_{12}\,P_{13}
\\ &
\qquad   \ +
  \ \frac{\rho_{12}+\rho_{23}-\rho_{13}}{m_2}\,P_{12}\,P_{23}\ + \
\frac{\rho_{23}+\rho_{13}-\rho_{12}}{m_3}\,P_{23}\,P_{13}
\ + \
\frac{\rho_{24}+\rho_{14}-\rho_{12}}{m_4}\,P_{24}\,P_{14}
\\ &
\qquad   \ + \
   \frac{\rho_{12}+\rho_{14}-\rho_{24}}{m_1}\,P_{12}\,P_{14} \ + \
   \frac{\rho_{12}+\rho_{24}-\rho_{14}}{m_2}\,P_{12}\,P_{24}\ + \
   \frac{\rho_{13}+\rho_{34}-\rho_{14}}{m_3}\,P_{13}\,P_{34}
   \\ &
\qquad   \ + \
   \frac{\rho_{14}+\rho_{34}-\rho_{13}}{m_4}\,P_{14}\,P_{34}  \ + \
   \frac{\rho_{13}+\rho_{14}-\rho_{34}}{m_1}\,P_{13}\,P_{14} \ + \
   \frac{\rho_{23}+\rho_{24}-\rho_{34}}{m_2}\,P_{23}\,P_{24}
   \\ &
\qquad   \ + \
   \frac{\rho_{23}+\rho_{34}-\rho_{24}}{m_3}\,P_{23}\,P_{34} \ + \
   \frac{\rho_{24}+\rho_{34}-\rho_{23}}{m_4}\,P_{24}\,P_{34}   \,\bigg] \ + \  V     \ .
\end{aligned}
\end{equation}

The Hamiltonian (\ref{Hredrho}) is written in what we call the $\rho-$representation (cf. (\ref{Hredzeroam})). Its kinetic energy is a linear polynomial function of the $\rho-$coordinates. Also, the cross terms in momentum variables associated with non-adjacent edges of the polytope of interaction are absent. In the free system ($V=0$), the Hamilton's equations for the momentum variables $P_{ij}$ do not depend on the $\rho-$variables at all.

\subsubsection{The metric $g^{\mu \nu}(\rho)$}

From (\ref{Hredrho}) we obtain the associated cometric
\begin{equation}
\label{gmn33-rho}
 g^{\mu \nu}(\rho)\ = \left(
\begin{array}{cccccc}
 \frac{2 \left(m_1+m_2\right) \rho _{12}}{m_1 m_2} & \frac{\rho _{12}+\rho _{13}-\rho _{23}}{m_1} & \frac{\rho _{12}+\rho _{14}-\rho _{24}}{m_1} & \frac{\rho _{12}-\rho _{13}+\rho _{23}}{m_2} & \frac{\rho _{12}-\rho _{14}+\rho _{24}}{m_2} & 0 \\
 \frac{\rho _{12}+\rho _{13}-\rho _{23}}{m_1} & \frac{2 \left(m_1+m_3\right) \rho _{13}}{m_1 m_3} & \frac{\rho _{13}+\rho _{14}-\rho _{34}}{m_1} & \frac{-\rho _{12}+\rho _{13}+\rho _{23}}{m_3} & 0 & \frac{\rho _{13}-\rho _{14}+\rho _{34}}{m_3} \\
 \frac{\rho _{12}+\rho _{14}-\rho _{24}}{m_1} & \frac{\rho _{13}+\rho _{14}-\rho _{34}}{m_1} & \frac{2 \left(m_1+m_4\right) \rho _{14}}{m_1 m_4} & 0 & \frac{-\rho _{12}+\rho _{14}+\rho _{24}}{m_4} & \frac{-\rho _{13}+\rho _{14}+\rho _{34}}{m_4} \\
 \frac{\rho _{12}-\rho _{13}+\rho _{23}}{m_2} & \frac{-\rho _{12}+\rho _{13}+\rho _{23}}{m_3} & 0 & \frac{2 \left(m_2+m_3\right) \rho _{23}}{m_2 m_3} & \frac{\rho _{23}+\rho _{24}-\rho _{34}}{m_2} & \frac{\rho _{23}-\rho _{24}+\rho _{34}}{m_3} \\
 \frac{\rho _{12}-\rho _{14}+\rho _{24}}{m_2} & 0 & \frac{-\rho _{12}+\rho _{14}+\rho _{24}}{m_4} & \frac{\rho _{23}+\rho _{24}-\rho _{34}}{m_2} & \frac{2 \left(m_2+m_4\right) \rho _{24}}{m_2 m_4} & \frac{-\rho _{23}+\rho _{24}+\rho _{34}}{m_4} \\
 0 & \frac{\rho _{13}-\rho _{14}+\rho _{34}}{m_3} & \frac{-\rho _{13}+\rho _{14}+\rho _{34}}{m_4} & \frac{\rho _{23}-\rho _{24}+\rho _{34}}{m_3} & \frac{-\rho _{23}+\rho _{24}+\rho _{34}}{m_4} & \frac{2 \left(m_3+m_4\right) \rho _{34}}{m_3 m_4} \\
\end{array}
\right) \ ,
\end{equation}
which is linear in $\rho$-coordinates, cf.(\ref{gmn33-red}). Its determinant reads
\[
D_m \ = \ \frac{576\,{(m_1+m_2+m_3+m_4)}^2}{ m_1^3\,m_2^3\,m_3^3\,{m_4^3}}\, {\cal V} \,(\,{\cal S}\,{\cal P} \ - \ 144\,m_1\,m_2\,m_3\,m_4\,{\cal V}\,) \ .
\]
Hence, $D_m$ is a polynomial function of the $\rho$-variables and it depends on three coordinates alone, namely the volume variables $\cal V$, $\cal S$ and $\cal P$. {Explicitly, the volume variables $\cal S$ and $\cal P$ in the $\rho-$representation read as follows:
\begin{equation}
\label{}
\begin{aligned}
{\cal S} \  = \ & m_1\,m_2\,m_3\,\big(2 \rho_{13} \,\rho_{12}+2 \rho_{23}\, \rho_{12}+2 \rho_{13}\, \rho_{23}-\rho_{13}^2-\rho_{23}^2-\rho_{12}^2  \big) \ + \
\\ &
m_1\,m_2\,m_4\,\big( 2 \rho_{14}\, \rho_{12}+2\rho_{24}\, \rho_{12}+2 \rho_{14}\, \rho_{24}-\rho_{12}^2-\rho_{14}^2-\rho_{24}^2\big) \ + \
\\ &
m_1\,m_3\,m_4\,\big(2 \rho_{14}\, \rho_{13}+2 \rho_{34}\, \rho_{13}+2 \rho_{14}\, \rho_{34}-\rho_{14}^2-\rho_{34}^2 -\rho_{13}^2\big) \ +\
\\ &
m_2\,m_3\,m_4\,\big( 2 \rho_{24}\, \rho_{23}+2 \rho_{34}\, \rho_{23}+2 \rho_{24}\, \rho_{34}-\rho_{24}^2-\rho_{34}^2-\rho_{23}^2 \big) \nonumber  \ ,
\end{aligned}
\end{equation}
\begin{equation}
\label{}
{\cal P} \ = \ \frac{m_1 m_2 \,\rho_{12}+m_1 m_3\, \rho_{13}+m_1 m_4 \,\rho_{14}+m_2 m_3\, \rho_{23}+m_2 m_4\, \rho_{24}+m_3 m_4\, \rho_{34}}{m_1+m_2+m_3+m_4} \nonumber  \ ,
\end{equation}
see (\ref{VVS})-(\ref{VVP}), and similarly for $\cal V$.}

The six dimensional $\rho$-space with cometric (\ref{gmn33-rho}) is not flat. The classical Hamiltonian ${H}_{\rho}$ (\ref{Hredrho}) is in complete agreement with the result obtained through the procedure of \emph{dequantization} from the corresponding quantum Hamiltonian operator.

\subsection{Four-body closed chain of interactive harmonic oscillators}

As an application of the $\rho-$representation, we consider the case
of a 4-body oscillator system. The defining potential $V^{(\text{har})}$ is given by
\begin{equation}
\label{V3-es}
   V^{(\text{har})}\ =\ 2\,\om^2\bigg[   \nu_{12}\,\rho_{12} \ + \  \nu_{13}\,\rho_{13} \ + \  \nu_{14}\,\rho_{14} \ + \ \nu_{23}\,\rho_{23} \ + \  \nu_{24}\,\rho_{24} \ + \  \nu_{34}\,\rho_{34} \bigg]\ ,
\end{equation}
being $\om > 0$ the frequency and $\nu_{12}, \nu_{13}, \nu_{14}, \nu_{23}, \nu_{24}, \nu_{34} \geq 0$ determine the spring constants. In general, the Hamiltonian (\ref{Hredrho}) with $V^{(\text{har})}$ does not admit separation of variables in $\rho-$coordinates. {However, unlike the $r$-representation, the potential is linear in $\rho-$variables. Upon a gauge rotation of the Hamiltonian, the corresponding $S-$states eigenfunctions are simply polynomial functions in $\rho-$variables. Also, for the simplest case of equal masses and equal spring constants the above potential $V^{(\text{har})}$ becomes a function of the volume variable $\cal P$ alone.}

Since the kinetic energy in the original Hamiltonian (\ref{Hamil}) decouples in Jacobi vectorial coordinates (\ref{Jacobi}), it is natural to look for separation of variables in such coordinates. A direct computation shows that, in general, the potential (\ref{V3-es}) is a non-diagonal quadratic expression in the Jacobi vectorial coordinates (\ref{Jacobi}). Nevertheless, the Principal Axes Theorem guarantees that it is always diagonalizable in certain variables. These variables are linear combinations of (\ref{Jacobi}) and the coefficients can be described by a rotation matrix (Euler angles) that preserves the kinetic energy term in diagonal form. The corresponding expressions are not particularly illuminating (rather cumbersome) and will not be displayed.

\textbf{Remark.} For arbitrary masses and frequencies, the 4-body oscillator system in $\mathbb{R}^d$ ($d>2$) can be transformed to a system of three $d-$dimensional decoupled isotropic harmonic oscillators where the frequencies, in general, are not commensurable. It implies that the original Hamiltonian (\ref{Hamil}) with potential $V^{(\text{har})}$ is an integrable system in full generality (even for non-zero values of angular momentum). In the case of 3-bodies with pairwise harmonic interactions at $d=3$, the integrability was demonstrated in Ref.\cite{Castro1993ExactSF} explicitly.

It can be shown that in the special case {of the three constraints imposed}
\begin{equation}
\label{super-int}
   m_1\,\nu_{34}\ =\ m_3\,\nu_{14}\quad ,  \qquad m_2\,\nu_{34}\ = \ m_3\,\nu_{24}\quad ,\qquad m_2\,\nu_{13}\ =\ m_1\,\nu_{23}\ ,
\end{equation}
the original Hamiltonian (\ref{Hamil}) with potential $V^{(\text{har})}$ reduces, in the center-of-mass reference frame, to the 3-body Jacobi oscillator system
\begin{equation}
\label{HJa}
   {\cal H}_J\ \equiv  \frac{1}{2}\bigg[ {({\bf p}^{(J)}_{1})}^2 \ + \ {({\bf p}^{(J)}_{2})}^2 \ + \ {({\bf p}^{(J)}_{3})}^2\,\bigg] \ + \ V^{(J)}\ ,
\end{equation}
\[
V^{(J)} \ = \   A_1\,{({r}^{(J)}_{1})}^2\ + \  A_2\,{({r}^{(J)}_{2})}^2\ + \  A_3\,{({r}^{(J)}_{3})}^2  \ ,
\]
${r}^{(J)}_{i}= \mid {\bf r}^{(J)}_{i} \mid$, $i=1,2,3$, see (\ref{Jacobi}); here we make the identifications
\begin{equation}
\begin{aligned}
& A_1 \ = \ \frac{2 \,\omega ^2\, \left(m_2 \,\nu _{12}+m_1\, \left(\nu _{12}+\nu _{23}+\nu _{24}\right)\ \right)}{m_1 \,m_2}\ , \quad A_2 \ = \ \frac{2 \,\omega ^2\, \left(m_1\, \nu _{23}+m_2\, \nu _{23}+m_3\, \left(\nu _{23}+\nu _{24}\right)\ \right)}{m_2\, m_3} \\ & \qquad A_3 \ = \ \frac{2\,\nu _{24} \,\omega ^2 \,\left(m_1+m_2+m_3+m_4\right) }{m_2\, m_4} \ .
\end{aligned}
\end{equation}
The symmetries of ${\cal H}_J$, which describes formally three $d-$dimensional isotropic oscillators (equivalently, an integrable anisotropic harmonic oscillator in the $3d-$dimensional space), were analyzed 
in \cite{Miguel A, Jauch} (and references therein). Now, if the additional constraint
\[
m_4\,\nu_{23}\ = \ m_3\,\nu_{24} \ ,
\]
is imposed in (\ref{HJa}), then two frequencies coincide $A_2 = A_3$. In this case, the 4-body Jacobi oscillator system as well as $V^{(\text{har})}$ becomes minimally superintegrable. Finally, the extra constraint
\[
m_3\,\nu_{12}\ = \ m_1\,\nu_{23} \ ,
\]
renders the system maximally superintegrable, all three {\it angular} frequencies coincide $A_1 = A_2 = A_3$. It may be mentioned that the quantum counterpart of the model was studied in detail in \cite{4bodyHarmonic}. Here, the relation of $V^{(\text{har})}$ to the 4-body Jacobi oscillator system clarifies and shows (in a simpler manner) the (super)integrability properties of the system. {Such an identification was not realized in Ref.\cite{4bodyHarmonic} being unknown in that time.}

\subsection{Generalized chain of interactive 4-body harmonic oscillators}

It is worth mentioning a generalization of the previous 4-body oscillator system (\ref{V3-es}). The corresponding potential $V^{(\text{har})}$, in the $\rho-$representation, reads
\begin{equation}
\label{V3-gen}
\begin{split}
   V_R^{(\text{har})}\  = \ & 2\,\om^2\bigg[   \nu_{12}\,{(\sqrt{\rho_{12}}-R_{12})}^2 \ + \  \nu_{13}\,{(\sqrt{\rho_{13}}-R_{13})}^2 \ + \  \nu_{14}\,{(\sqrt{\rho_{14}}-R_{14})}^2
\\ &    + \ \nu_{23}\,{(\sqrt{\rho_{23}}-R_{23})}^2 \ + \  \nu_{24}\,{(\sqrt{\rho_{24}}-R_{24})}^2 \ + \  \nu_{34}\,{(\sqrt{\rho_{34}}-R_{34})}^2 \bigg]\ ,
\end{split}
\end{equation}
where $R_{ij}>0$ play the role of rest lengths. We call it {generalized harmonic oscillator system
or 4-body harmonic molecule}. Especially, the  tetrahedron of interaction with sides $R_{ij}$ defines equilibrium configuration. In this case, the Hamiltonian (\ref{Hredrho}) with potential $V_R^{(\text{har})}$ does not admit a separation of variables neither in $\rho-$coordinates nor in other coordinate system. In the simpler 3-body case on the plane $d=2$, identical particles and equal spring and rest lengths constants, the corresponding generalized harmonic system exhibits a rich dynamics as a function of the values of parameters and the energy. For instance, a power-law statistics, that fits the Levi-walk model \cite{Zaburdaev}, develops \cite{Saporta Katz, Katz}. Recently, an experimental realization of this 3-body system was assembled using analog electrical elements \cite{Classicalcase}.

{{By substituting  $\rho_{12}={(R + A_{12}\,e^{i\,\Omega\,t})}^2$, $\rho_{13}={(R + A_{13}\,e^{i\,\Omega\,t})}^2$, $\rho_{23}={(R + A_{23}\,e^{i\,\Omega\,t})}^2$ into the Newton's equations of motion for the Hamiltonian (\ref{Hredrho}) with potential $V_R^{(\text{har})}$ (\ref{V3-gen}) and requiring all the linear terms in the constants $A_{ij}$ to vanish}, it can be shown that in the regime of small oscillations ({$A_{ij}\ll 1$}), the symmetric 4-body harmonic oscillator system with equal masses $m_i=1$, equal spring constants $\nu_{ij}=1$ and equal rest lengths $R_{ij}=R>0$ 
($i\neq j=1,2,3,4$) possesses 6 normal modes with frequencies $\Omega=\sqrt{8}\,\omega$ (symmetric stretch), $\Omega=\sqrt{\frac{8}{5}}\,\omega$ (triple degenerate) and $\Omega=\sqrt{\frac{8}{7}}\,\omega$ (doubly degenerate), respectively. The symmetric stretch represents an exact periodic trajectory of the system beyond small oscillations.}

\section{Volume variables representation ($d>2$)}

In this Section we consider the special family of potentials $V$ that can be written in terms of the volume variables ${\cal V}$, $\cal S$ and $\cal P$ defined in (\ref{VVV}), (\ref{VVS}), (\ref{VVP}), respectively. Namely,
\[
V \ = \ V({\cal V},\,{\cal S},\,{\cal P}) \ .
\]
Let us recall that variable $\cal V$ is the square of the volume of the tetrahedron of interaction, $\cal S$ is the weighted sum of the 4 areas (squared) of the faces (triangles) of the tetrahedron whilst $\cal P$ is the weighted sum of the 6 edges (squared) of the tetrahedron. Clearly, $\cal P$ is an homogeneous polynomial of degree 1 in $\rho$-variables whereas $\cal S$ and $\cal V$ are polynomials of degree 2 and 3, respectively.

\noindent
\textbf{Remark.} Each of these three volume variables is $S_4$ invariant under the permutation of any pair of the four particles, volume variables remain unchanged under the permutation $({\bf r}_{i},\,m_i) \Leftrightarrow ({\bf r}_{j},\,m_j)$. {However, even in the simplest case of equal masses $m_1=m_2=m_3=m_4$, the variables $\cal V$ and $\cal S$ are not $S_6$ invariant under the permutation of any pair of the six $\rho-$coordinates.}

Explicitly, in the reduced 12-dimensional phase space $\Gamma_r$ of relative motion, we perform the following change of variables
\begin{equation}
\label{CC5b}
\{r_{12},\,r_{13},\,r_{14},\,r_{23},\,r_{24},\,r_{34}\}\ \Rightarrow \
\{  {\cal V},\,{\cal S},\,{\cal P},\,\varphi_1,\,\varphi_2,\,\varphi_{3}  \} \ ,
\end{equation}

here the generalized coordinates $\varphi_i$ ($i=1,2,3$) denote any set of variables such that the transformation (\ref{CC5b}) is canonical. Again, if the potential $V$ does not depend on the $\varphi-$variables, it follows that the hypersurface
\begin{equation}
\label{IC5b}
\Gamma_{\rm vol} \ \equiv \ \Gamma_{r}|_{\ p_{_{\varphi_1}} =\,p_{_{\varphi_2}}=\,p_{_{\varphi_{3}}}=0} \quad ; \,\qquad  \Gamma_{\rm vol} \ \subset \ \Gamma_{r}\ \subset \ \Gamma_{\rm rel}
\end{equation}
becomes an invariant manifold\cite{Wiggins}.

Thus, by taking the restriction of ${H}_{r}$ on $\Gamma_{\rm vol}$ we obtain a further effective Hamiltonian (${ H}_{r} \rightarrow {\cal H}_{\rm vol}$).

As a result of standard computation the Hamiltonian that governs the trajectories that solely depend on the three volume coordinates is of the form
\begin{equation}
\label{HVV}
\begin{aligned}
{\cal H}_{\rm vol} \ = \ & \frac{1}{72\,m}{\cal V}\,{\cal S}\,P^2_{{\cal V}} \ + \ \big(  3456\,m\,M\,{\cal V} \,+\, 8\,M\,{\cal S}\,{\cal P}   \big)\,P^2_{{\cal S}} \ + \ 2\,{\cal P}\,P^2_{{\cal P}}
\\ & + \ 32\,M\,{\cal V}\,{\cal P}\,P_{\cal V}\,\,P_{\cal S} \ + \ 12\,{\cal V}\,P_{\cal V}\,\,P_{\cal P} \ + \ 8\,{\cal S}\,P_{\cal S}\,P_{\cal P} \ + \ V({\cal V},\,{\cal S},\,{\cal P})\ ,
\end{aligned}
\end{equation}
here $m \equiv m_1\,m_2\,m_3\,m_4$, $M \equiv m_1+m_2+m_3+m_4$ and $P_{\cal V}$,\,$P_{\cal S}$, $P_{\cal P}$ are the canonical momentum variables {conjugate to coordinates ${\cal V,S,P}$, respectively.} The Hamiltonian (\ref{HVV}) describes a three-dimensional particle moving in a curved space. Tensor of inertia can be identified with cometric in this case as well.
The associated cometric for (\ref{HVV}) is given by
\begin{equation}
\label{gmn33-geo}
 g_{}^{\mu \nu}({\cal V},\,{\cal S},\,{\cal P})\ = \left|
\begin{array}{ccc}
\frac{1}{72\,m}{\cal V}\,{\cal S}  & \ 16\,M\,{\cal V}\,{\cal P} \ & \ 6\,{\cal V} \\
            &                                   &          \\
  16\,M\,{\cal V}\,{\cal P} & \  3456\,m\,M\,{\cal V} \,+\, 8\,M\,{\cal S}\,{\cal P}  \ & \ 4\,{\cal S} \\
            &  \                                &          \\
 6\,{\cal V}\ & \ 4\,{\cal S}\ &\ 2\,{\cal P}
\end{array}
               \right| \ .
\end{equation}
Its components are polynomials in the volume variables. Its determinant admits factorization,
\begin{equation}
\label{DetVV}
  \text{Det}\big[\,g_{}^{\mu \nu}({\cal V},\,{\cal S},\,{\cal P})\,\big]\ =\ -\frac{2 {\cal V} \left(559872 m^2 M {\cal V}^2+2304 m M^2 {\cal P}^3 {\cal V}-2592 m M {\cal P} {\cal S} {\cal V}-M {\cal P}^2 {\cal S}^2+{\cal S}^3\right)}{9\, m} \ .
\end{equation}
This determinant vanishes at
\[
{\cal V} \ = \ 0 \ ,
\]
i.e. when the four particles move on a plane or are collinear. Moreover, the second factor in (\ref{DetVV}) also vanishes for certain configurations with non-zero $\cal V$. {In particular, if all four masses are equal then $\cal V$ vanishes for {the configuration formed by an equilateral triangle base and three equal isosceles triangle sides as well as for} a regular tetrahedron. The space with cometric (\ref{gmn33-geo}) is not flat.}

\vspace{0.2cm}

\subsection{Equations of motion}

\vspace{0.2cm}

From the Hamiltonian (\ref{HVV}), we obtain the following Hamilton's equations of motion:

\begin{equation}
\label{EQMT}
\begin{aligned}
& \dot P_{\cal P} \ = \ -2\,(\, P_{\cal P}^2 \ + \ 4\, M\, P_{\cal S}(\,P_{\cal S}\,S \ + \ 4\,P_{\cal V}\,{\cal V} ) \, ) \ - \ \partial_{\cal P}V
\\ &
 \dot P_{\cal S} \ = \ -8\,P_{\cal P}\,P_{\cal S} \ - \ 8\,M\,{\cal P}\,P_{\cal S}^2 \ -  \ \frac{{\cal V}\,P_{\cal V}^2}{72\,m} \ - \ \partial_{\cal S}V
\\ &
 \dot P_{\cal V} \ = \ -3456\,m\,M\,P_{\cal S}^2 \ - \ 4\,(\, 3\,P_{\cal P} \ + \ 8\,M\,{\cal P}\,P_{\cal S}  \,)\,P_{\cal V} \ -  \ \frac{{\cal S}\,P_{\cal V}^2}{72\,m} \ - \ \partial_{\cal V}V
\\ &
 \dot {\cal P} \ = \ 4\,(\, {\cal P}\,P_{\cal P} \ + \ 2\,{\cal S}\,P_{\cal S} \ + \ 3\,{\cal V}\,P_{\cal V}  \,)
\\ &
 \dot {\cal S} \ = \ 8\,(\,{\cal S}\,P_{\cal P} \ + \ 2\,M\,( {\cal P}\,{\cal S}\,P_{\cal S} \ + \ 432\,m\,{\cal V}\,P_{\cal S}  ) \ + \ 2\,{\cal P}\,{\cal V}\,P_{\cal V}   \,)
\\ &
 \dot {\cal V} \ = \ 12\,{\cal V}\,P_{\cal P} \ + \ 32\,M\,{\cal P}\,{\cal V}\,P_{\cal S} \ + \ \frac{{\cal S}\,{\cal V}\,P_{\cal V}}{36\,m}
\\ &
\end{aligned}
\end{equation}

\subsection{Two-variable potentials $V  =  V({\cal S},\,{\cal P})$: planar case $d=2$}

For the 2-variable potentials of the form
\begin{equation}\label{PVS}
V \ = \ V({\cal S},\,{\cal P}) \ ,
\end{equation}
the above equations of motion (\ref{EQMT}) admit particular special solutions with ${\cal V} =0$ where the problem becomes effectively two-dimensional (2 degrees of freedom). The corresponding reduced Hamiltonian is given by
\begin{equation}
\label{HVV2D}
\begin{aligned}
{h}_{\rm vol} \ =  \ &   8\,M\,{\cal S}\,{\cal P} \,P^2_{{\cal S}} \ + \ 2\,{\cal P}\,P^2_{{\cal P}}
\ + \ 8\,{\cal S}\,P_{\cal S}\,P_{\cal P} \ + \ V({\cal S},\,{\cal P})\ .
\end{aligned}
\end{equation}

The condition ${\cal V}=0$ implies that the volume of the tetrahedron of interaction vanishes. On the plane $d=2$, this is the case  necessarily. We call (\ref{HVV2D}) the (${\cal P},\,{\cal S}$)-representation.

\subsubsection{The potential $V=\frac{{U[P/ \sqrt{S}]}}{\sqrt{S}}$}

Now, within the (${\cal P},\,{\cal S}$)-representation we consider the physically relevant case of
\[
\dot {\cal P}\ = \ 0 \ ,
\]
namely, a system with constant moment of inertia.

Specifically, we consider an arbitrary potential $V=V({\cal P},{\cal S})$ on the submanifold of phase space for which ${h}_{\rm vol}=0$ (zero energy level) and $\dot{{\cal P}}=0$ (a constant moment of inertia). In this case, from (\ref{EQMT}) it follows that the potential must obey the following linear partial differential equation
\begin{equation}\label{PMG}
2\,{\cal S}\,\partial_{\cal S}\,V \ + \ {\cal P}\,\partial_{\cal P}\,V \ = \ -V \ .
\end{equation}
The solution of (\ref{PMG}) is given by
\begin{equation}\label{VPconst}
V\ =\ \frac{{U[{\cal P}/ \sqrt{{\cal S}}]}}{\sqrt{{\cal S}}} \ ,
\end{equation}
where $U[z]$ is an arbitrary function of the argument $z$.

\vspace{0.2cm}

For an arbitrary function $U[{\cal P}/ \sqrt{{\cal S}}]$ we obtain the first-order nonlinear ordinary differential equation for the volume variable $S$
\begin{equation}\label{}
{\cal P}\,{\dot{{\cal S}}}^2 \ - \ 32\,U\,(M\,{\cal P}^2 \ - \ {\cal S})\,\sqrt{{\cal S}} \ = \ 0 \ ,
\end{equation}
here ${\cal P}>0$ plays the role of an external parameter. For the potential $V=-\gamma \frac{{\cal P}}{{\cal S}}$ we arrive to the solution
\begin{equation}\label{}
  {\cal S}(t) \ = \  M \,{\cal P}^2\ +\ \frac{1}{4} \,\gamma \, \left(\,32\, t^2\,-\,8\, \sqrt{2} \,c_1\, t\,+\,c_1^2\right) \ > \ 0 \ ,
\end{equation}
$c_1$ is a constant of integration. Notice that the potential $V \propto \frac{1}{{\cal P}}$ leads to the solutions ${\cal S}(t)=M\,{\cal P}^2$ (for four equal unit masses $M=4$, hence, a regular tetrahedron) and ${\cal S}(t)=0$ (four particles on a line).

As a function of the two lowest order volume variables, those that are linear and quadratic expressions in $\rho-$variables, respectively, the same potential $V=-\gamma \frac{{\cal P}}{{\cal S}}$ occurs in the three-body case. It was originally studied
in Ref.\cite{Montgomery} where the author builds the hyperbolic plane with its geodesic flow as the scale plus symmetry reduction of a 3-body problem in the Euclidean plane.

\subsubsection{The metric $g_{\rm vol}^{\mu \nu}$}

Now, the metric (or, equivalently, the tensor of inertia) for (\ref{HVV2D}) is of the form
\begin{equation}
\label{VGemVol}
 g_{\rm vol}^{\mu \nu}\ = \left|
 \begin{array}{cc}
 2\,{{\cal P}} & \ 4\,{\cal S}  \\
  4\,{\cal S} & 8\,M\,{\cal P}\,{\cal S}     \,
 \end{array}               \right| \ ,
\end{equation}
with determinant
\[
  D_{\rm vol}\ = \ 16\,{\cal S}\,(\,M\,{{\cal P}}^2 \ - \ {\cal S}\,)\ .
\]
In the case of a \emph{regular tetrahedron} with four equal unit masses the relation $4\,{{\cal P}}^2 \ = \ {\cal S}$ holds and this determinant vanishes, ${D}_{\rm vol}=0$.
It also follows from (\ref{VGemVol}) that the corresponding Ricci scalar $\rm {Rs}$ is given by
\begin{equation}
\label{Rs1}
 {\rm  Rs}_{\rm vol} \ = \   \frac{32\,M\,{\cal P}\,{\cal S}\,(\, 4\,M\,{\cal S} \ - \ 1 \,)}
     {{D_{\rm vol}}^2} \ .
\end{equation}

\bigskip

\subsection{One-variable potentials: $V =  V({\cal P})$ }

{
An interesting particular case occurs when the potential (\ref{PVS}) does not depend on $\cal S$ and is a function of the volume variable $\cal P$ alone
\[
V  \ = \ V({\cal P}) \ .
\]
It follows from (\ref{HVV2D}) that there exist trajectories which depend on $\cal P$ only. Such trajectories lie on the intersection between the hypersurfaces ${\cal V}=0$ and $P_{\cal S}=0$, they are described by the Hamiltonian
\begin{equation}
\label{HPP}
 {h}_{\rm vol}\mid_{{}_{\small P_{\cal S}=0}}\ \equiv \  {\cal H}_{\cal P} \ =\ 2\,{\cal P}\,P_{\cal P}^2\ +\ V({\cal P})\ ,
\end{equation}
The connection between the trajectories of ${\cal H}_{\cal P}$ (\ref{HPP}) with those of ${h}_{\rm vol}$ (\ref{HVV2D}) is the following. Taking a non-trivial solution ${\cal P}={\cal P}(t)\neq 0$ of (\ref{HPP}) one can define the function ${\cal S}={\cal S}(t)$ for which the canonical momentum $P_{\cal S}$ vanishes in ${h}_{\rm vol}$. Explicitly, the condition can be obtained from (\ref{HVV2D}),
\begin{equation}
\label{PSzero}
  P_{\cal S} \ = \   \frac{2 \,{\dot P} \,{\cal S} \,-\,P\,{\dot {\cal S}}}{16\,{\cal S}\,( {\cal S}\,-\,M\,{\cal P}^2)} \ = \ 0 \ .
\end{equation}
Such ${\cal P}={\cal P}(t)$ and ${\cal S}={\cal S}(t)=\lambda\,{\cal P}^2(t)$, with $\lambda \neq M$ a non-negative parameter, satisfy the equations of motion for ${h}_{\rm vol}$. Moreover, they also obey the equations (\ref{EQMT}) for ${\cal H}_{\rm vol}$ with ${\cal V}=0$, hence, the system degenerates into a planar system. At $\lambda=0$, thus ${\cal S}=0$, the problem further degenerates to the line.

\subsubsection{Anharmonic Oscillator potential}

As a concrete example, let us consider the following one-variable potential
\begin{equation}
\label{VANP}
\begin{aligned}
  V^{(AO)} & \ = \ A\,{\cal P} \ + \ B \,{\cal P}^2\ =
\\ &
 A\,\frac{m_1 m_2 \,\rho_{12}+m_1 m_3 \,\rho_{13}+m_1 m_4 \,\rho_{14}+m_2 m_3 \,\rho_{23}+m_2 m_4 \,\rho_{24}+m_3 m_4 \,\rho_{34}}{M}
\\ &
\ +
B\,{\bigg[\frac{m_1 m_2 \,\rho_{12}+m_1 m_3 \,\rho_{13}+m_1 m_4 \,\rho_{14}+m_2 m_3 \,\rho_{23}+m_2 m_4 \,\rho_{24}+m_3 m_4 \,\rho_{34}}{M}\bigg]}^2  \ ,
\end{aligned}
\end{equation}

\noindent
where (i) $B \geq 0$ and (ii) if $B=0$ the parameter $A > 0$. In terms of variable $P$ the potential (\ref{VANP}) corresponds to a shifted one-dimensional harmonic oscillator on the half line $\mathbb{R}_+$ with non-standard kinetic energy {$2\,{{\cal P}}\,P_{{\cal P}}^2$, see below}. In the $\rho-$representation it corresponds to a certain six-dimensional quadratic potential, defined on a certain subspace of $\mathbb{R}^6_+$, whilst in the $r-$representation it describes a six-dimensional anisotropic harmonic oscillator with the quartic anharmonicity.

For the potential (\ref{VANP}), the Hamiltonian (\ref{HPP}) becomes
\begin{equation}
\label{HPPAO}
\begin{aligned}
  {\cal H}^{(AO)}_{\cal P} \  & = \ 2\,{{\cal P}}\,P_{{\cal P}}^2  \ + \ A\,{\cal P} \ + \ B \,{\cal P}^2  \ .
\end{aligned}
\end{equation}

The time-evolution of the system is given by
\begin{equation}
\label{PtSolu}
{\cal P}(t) \ =\ x^2\ = \ \frac{A\,k^2}{B\,(1-k^2)}\,\text{sn}^2(y,ik) \quad ,\qquad y\,=\, \pm \sqrt{\frac{2\,A}{1-k^2}}\ t\ ,
\end{equation}
($B \neq 0$, $k \neq \pm 1$) where sn$(y,ik)$ is the Jacobi elliptic function with imaginary elliptic modulus $(ik)$. In the case of physical systems the function $P(t)$ should be positive and the parameter $k \neq \pm 1$ is real.

For the solution (\ref{PtSolu}) the Hamiltonian has a meaning of energy and it takes the value
\[
{\cal H}^{(AO)}_P\ = \ \frac{A^2\, k^2}{B \left(1-k^2\right)^2}\ .
\]

Putting the $B = 0$ in (\ref{HPPAO}) we obtain the harmonic oscillator potential $V=A\,{\cal P}$, which is quadratic polynomial in the $r$-representation. In this case the trajectories are trigonometric vibrations, ${\cal P}(t)  =   c_1\,\cos^2(\sqrt{9\,A/2}\,\,t+c_2)$ with energy ${\cal H}^{(AO)}_{\cal P}\, =\, \ A\,c_1$ where $c_1>0, c_2$ are real constants of integration. In general, due to the energy conservation ${\cal H}^{(AO)}_{\cal P}=E$ the trajectories in the phase space are always cubic curves,
\[
   2\,{{\cal P}}\,P_{{\cal P}}^2  \ + \ A\,{\cal P} \ + \ B \,{\cal P}^2 \ =\ E\ .
\]
Note that at fixed values of the energy $E$ and $A$ the presence of anharmonicity ($B>0$) in (\ref{HPPAO}) tends to decrease the amplitude of the harmonic motion $(B=0)$. Also, the harmonic potential $V=A\,{\cal P}$ corresponds to the aforementioned 4-body chain of harmonic oscillators. From the definition of $\cal P$, it follows that it can be written as a 4-body maximally superintegrable Jacobi oscillator as well.

\section{u-variables representation ($d>1$)}

In this Section we consider 4 identical particles, $m_1=m_2=m_3=m_4=1$, and focus on the three-variable family of potentials $V$ of the form,
\[
V \ = \ V(u_1,\,u_2,\,u_3) \ ,
\]
where
\begin{equation}
u_1 \ = \ \rho_{12} \ + \  \rho_{34} \quad , \qquad u_2 \ = \ \rho_{13} \ + \  \rho_{24} \quad , \qquad u_3 \ = \ \rho_{14} \ + \  \rho_{23}  \ ,
\end{equation}

\noindent
are linear functions of the $\rho-$variables and each of them corresponds to the sum of two disconnected edges squared of the tetrahedron of interaction. Clearly, they also are $S_4-$permutationally invariant under the permutations of any pair of bodies. The sum of all three $u-$variables is given by the volume variable ${\cal P}=u_1+u_2+u_3$.

The Hamiltonian that governs the trajectories that solely depend on the three $u-$coordinates reads

\begin{equation}
\label{Huu}
\begin{aligned}
{\cal H}_{\rm u} \ = \ & 4\,(\,u_1\,P_{u_1} \ + \ u_2\,P_{u_2}\ + \ u_3 \,P_{u_3}\,) \ + \ 4\,(u_1 \,+\, u_2\, - \, u_3)\,P_{u_1}\,P_{u_2}
\\ & \ + \ 4\,(u_1 \,+\, u_3\, - \, u_2)\,P_{u_1}\,P_{u_3}\ + \ 4\,(u_2 \,+\, u_3\, - \, u_1)\,P_{u_2}\,P_{u_3}
\ + \ V(u_1,\,u_2,\,u_2)\ ,
\end{aligned}
\end{equation}
here $P_{u_2}$,\,$P_{u_2}$, and $P_{u_3}$ are the corresponding canonical momentum variables. The Hamiltonian (\ref{Huu}) describes a three-dimensional particle moving in a curved space. Tensor of inertia can be identified with cometric in this case as well. Unlike the volume variables, for the planar system $d=2$ all three $u-$variables are, in general, always different from zero.

\section{Conclusions}
\label{Conclusion}

In the present study the four-body classical system of four arbitrary masses in Euclidean space $\mathbb{R}^{4d}$ as the coordinate space is considered. We assume the generic potentials $V=V(r_{ij})$, not necessarily pairwise interactions, depending on the 6 relative mutual distances $r_{ij}$ between bodies are considered.

For trajectories of the total angular momentum equal to zero the original $4d-$dimensional problem is reduced to a six-dimensional problem defined in the space of relative (internal) radial motion. It realizes a symmetry reduction from $(4d)$-dimensional coordinate space to six-dimensional space of relative distances.
The corresponding reduced Hamiltonian is constructed explicitly using two representations: (I) the \textit{$r-$representation} where the generalized coordinates are the six relative distances $r_{ij}$ and where the kinetic energy term (free Hamiltonian) is a second order polynomial in the associated canonical momenta $p_{ij}$ with rational coefficients in the variables $r_{ij}$, and (II) the \textit{$\rho-$representation} where the free Hamiltonian is a polynomial function of both the $\rho-$variables, $\rho_{ij}=r_{ij}^2$, and their canonical momenta $P_{ij}$. The cometric appearing in} the corresponding Hamiltonians $H_r$ (\ref{Hredzeroam}) and
$H_\rho$ (\ref{Hredrho}), respectively, describes a six-dimensional particle moving in a curved space.

As a particular example, the 4-body harmonic oscillator system was studied within the $\rho-$representation in detail. {Moreover, in Jacobi vectorial variables due to the separation of variables}, the original Hamiltonian can be reduced to a 3-body Jacobi oscillator integrable system, {where the potential is the sum of squares of Jacobi vectorial coordinates.} {We emphasize that this property is exclusive of the Jacobi variables and is valid even for a non-zero value of the angular momentum.}  Furthermore, for specific relations between the masses and the frequencies occurring in the harmonic oscillator potential, the system becomes superintegrable. A non-integrable generalization (4-body harmonic molecule) of this model was briefly analyzed in the regime of small oscillations.

At $d>2$ the positions of the 4 bodies form a Platonic solid, the tetrahedron. This geometric object naturally defines the volume variables ${\cal V}$ (\ref{VVV}), $\cal S$ (\ref{VVS}) and $\cal P$ (\ref{VVP}). In fact, the variable $\cal P$ coincides with the moment of inertia of the system. Moreover, in the case of four equal masses $m_1=m_2=m_3=m_4$ each of the volume variables is $\mathcal{S}_4$-invariant under the interchange of any pair of the particles (vertices) in the tetrahedron.

Assuming $d>2$ and the potential depends on three volume variables $V=V({\cal V},\,{\cal S},\,{\cal P})$, the class of trajectories that depend only on the volume variables is governed by a six-dimensional Hamiltonian ${\cal H}_{\rm vol}$ (\ref{HVV}) in the phase space. We call it \emph{volume-representation of the Hamiltonian.} It describes a three-dimensional particle moving in a space which is non-flat. For $d = 2$, where ${\cal V}=0$, we indicate a generic two-variable potential $V=V({\cal S},\,{\cal P})$, for which the moment of inertia ${\cal P}$ is an integral of motion. Also, a further reduction can be performed in the case $V=V({\cal P})$. Here, the system admits a family of special ${\cal P}-$dependable trajectories, it described by a certain two-dimensional reduced Hamiltonian in the phase space which is derived.

For the case of identical particles with equal spring constants, the special $u-$\textit{representation} was developed, if the potential depends only on the $u-$variables (each of them being the sum of two disconnected edges squared of the tetrahedron of interaction). Each of these three coordinates is $\mathcal{S}_4$-invariant under the permutation of any pair of the particles (vertices) in the tetrahedron. They are evidently related with the volume variable ${\cal P} = u_1+u_2+u_3$.

The presented 4-body studies at $n=4$ together with previous 3-body ones at $n=3$, reported in \cite{IJMP1}, 
lead us to a certain conjectures for the general $n$-body case in $\mathbb{R}^d$ at $d > n-2$:

\begin{itemize}

\item The $n-$body harmonic oscillator potential
\[
V^{(\rm har)}_n \ = \ \omega^2\,\sum_{i>j}^n \nu_{ij}\,r_{ij}^2 \ = \ 
\omega^2\,\sum_{i>j}^n \nu_{ij}\,\rho_{ij} \quad , \qquad \nu_{ij} > 0 \ ,
\]
is integrable, it admits a complete separation of variables when they are taken as linear combinations 
of the Jacobi vectorial coordinates. The original Hamiltonian, in the center-of-mass reference frame, corresponds to  $(n-1)-$variables Jacobi oscillator system. By imposing the conditions on the masses 
and string constants $\frac{m_i}{m_j}=\frac{\nu_{ik}}{\nu_{jk}}$ ($i\neq j\neq k$), converts the system 
into maximally superintegrable. In this case, the harmonic potential has the form 
$V^{(\rm har)}_n \propto {\cal P}_n$, hence, it is proportional to the moment of inertia of the system 
$${\cal P}_n\ =\ \frac{\sum_{i>j}m_i\,m_j\,r_{ij}^2}{\sum_{i}m_i}\ .$$

\item At fixed $n$, the polytope of interaction (formed by taking bodies positions as vertices) defines $(n-1)$ volume variables. There exist trajectories,
that depend solely on these volume variables, the corresponding (reduced) Hamiltonian is a polynomial function.

\item For $d=2$, the potential of the form

\[
V({\cal P}_n,\,{\cal S}_n) \ = \ \frac{ U[{\cal P}_n/{\cal S}_n]} { \sqrt{{\cal S}_n} } \ ,
\]

where ${\cal S}_n$ is the weighted sum of the squares of the area of the faces (triangles) 
of the degenerate polytope of interaction and ${\cal P}_n$ is the sum squared of lengths of 
its edges, and $U[z]$ is an arbitrary function of $z$, possesses an infinite family 
of planar trajectories for which the variable ${\cal P}_n$ is an integral of motion.

\item At any dimension $d$ the potential $V= A\,{\cal P}_n \,+\,B\,{\cal P}_n^2$  
(a type of the quartic anharmonic potential in $r-$variables), admits particular solutions 
where
\begin{equation}
{\cal P}_n(t) \ \propto \ \frac{A\,k^2}{B\,(1-k^2)}\,\text{sn}^2(y,ik) \quad ,\qquad y\,=\, \pm \sqrt{\frac{A}{1-k^2}}\ t\ ,
\end{equation}
($A,B \neq 0$, $k \neq \pm 1$) is given by a Jacobi elliptic function with imaginary elliptic 
modulus $(ik)$.

\end{itemize}

At $d \leq n-2$, as for example 3 bodies on a line or 4 bodies on the plane, the polytope of interaction is degenerate. Thus, not all the $\frac{n(n-1)}{2}$ $r$-variables are independent and the polynomiality of the $L=0$ reduced Hamiltonian in $\rho-$representation is absent. In this case, the number of non-vanishing volume-variables is equal to $d$.   

\section{Author's contributions}

All authors contributed equally to this work.

\section{Acknowledgments}

We thank Willard Miller for useful discussions in the early stages of this study. This work was supported partially by CONACyT grant A1-S-17364 and DGAPA grant
IN113022 (Mexico). A.V.T. thanks PASPA-UNAM for a support during his sabbatical
stay at University of Miami, where this work was completed.

\section{DATA AVAILABILITY}

Data sharing is not applicable to this article as no new data were created or analyzed in this
study.


\end{document}